# THE DISCOVERY OF QUASARS AND ITS AFTERMATH

**K.I. Kellermann**

*National Radio Astronomy Observatory, 520 Edgemont Road,
Charlottesville, VA 22901, USA.*
Email: kkellerm@nrao.edu

**Abstract:** Although the extragalactic nature of quasars was discussed as early as 1960 by John Bolton and others it was rejected largely because of preconceived ideas about what appeared to be an unrealistically-high radio and optical luminosity. Following the 1962 observation of the occultations of the strong radio source 3C 273 with the Parkes Radio Telescope and the subsequent identification by Maarten Schmidt of an apparent stellar object, Schmidt recognized that the simple hydrogen line Balmer series spectrum implied a redshift of 0.16. Successive radio and optical measurements quickly led to the identification of other quasars with increasingly-large redshifts and the general, although for some decades not universal, acceptance of quasars as being by far the most distant and the most luminous objects in the Universe. However, due to an error in the calculation of the radio position, it appears that the occultation position played no direct role in the identification of 3C 273, although it was the existence of a claimed accurate occultation position that motivated Schmidt's 200-in Palomar telescope investigation and his determination of the redshift.

Curiously, 3C 273, which is one of the strongest extragalactic sources in the sky, was first catalogued in 1959, and the 13$^{th}$ magnitude optical counterpart was observed at least as early as 1887. Since 1960, much fainter optical counterparts were being routinely identified, using accurate radio interferometer positions which were measured primarily at the Caltech Owens Valley Radio Observatory. However, 3C 273 eluded identification until the series of lunar occultation observations led by Cyril Hazard. Although an accurate radio position had been obtained earlier with the Owens Valley Interferometer, inexplicably 3C 273 was misidentified with a faint galaxy located about one arc minute away from the true position. It appears that the Parkes occultation position played only an indirect role in the identification of the previously-suspected galactic star, which was only recognized as the optical counterpart after Schmidt's 200-in observations showed it to have a peculiar spectrum corresponding to a surprisingly-large redshift.

## 1 HISTORICAL BACKGROUND

The discovery of quasars in 1963, and more generally, active galactic nuclei (AGN), revolutionized extragalactic astronomy. In early February 1963, Maarten Schmidt (b. 1929; Figure 1) at Caltech recognized that the spectrum of the 13$^{th}$ magnitude apparently stellar object identified with the radio source 3C 273 could be most easily interpreted by a redshift of 0.16. Subsequent work by Schmidt and others led to increasingly-large measured redshifts and the recognition of the broad class of active galactic nuclei (AGN) of which quasars occupy the high luminosity end. Schmidt's discovery changed extragalactic astronomy in a fundamental way. Within a few years redshifts as great as 2 or more were being routinely observed, making possible a new range of cosmological studies as well as the realization that supermassive black holes which power radio galaxies and quasars play a prominent role in the evolution of galaxies. But the path to this understanding was a slow, tortuous one, with missed turns that could have, and should have, earlier defined the nature of quasars.

The events leading up to the recognition of quasars as the extremely luminous nuclei of distant galaxies go back much earlier than 1963; indeed, one wonders why the extragalactic nature of quasars was not recognized well before 1963, and why 3C 273, which is the seventh brightest radio source in the northern sky away from the Galactic Plane, was not identified at least one or two years earlier based on the radio position determined by observations carried out at the Owens Valley Radio Observatory (henceforth OVRO), which was more accurate than the occultation position used by Schmidt to identify 3C 273 in December 1962.

In the remainder of this section we review the early arguments and evidence for powerful activity in the nuclei of galaxies. In Section 2, we briefly review the status of extragalactic radio astronomy prior to the identification of 3C 48, and in Section 3, the identification of 3C 48, which might have been the first discovered quasar, but was unrecognized as such until the work on 3C 273 more than two years later, as described in Section 4. In Section 5 we return to the issues surrounding 3C 48, and in Sections 6 and 7 the implications for cosmology and the arguments for non-cosmological interpretations of quasar redshifts. Sections 8 and 9 describe the discovery of radio-quiet quasars. Finally, in Section 10 we summarize the history and highlight remaining issues and questions surrounding the discovery of quasars.

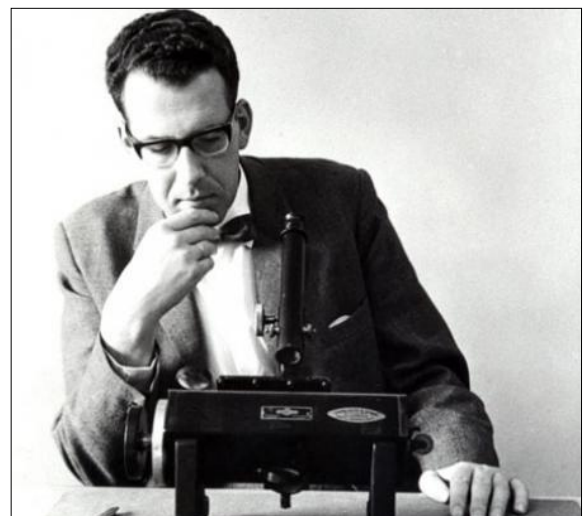

Figure 1: A photograph of Maarten Schmidt at work in 1965 (credit: James McClanahan, Engineering and Science, May, 1965; courtesy: Caltech Archives).

Probably the first person to note enhanced activity in the nucleus of a galaxy was Edward Arthur Fath (1880–1959) who reported on the nuclear emission line spectrum of NGC 1068 (Fath, 1909). Later observations of strong nucle-ar emission lines by Vesto Melvin Slipher (1875–1969) in 1917, Edwin Powell Hubble (1889–1953) in 1926, Milton Lasell Humason (1891–1972) in 1932, and Nicholas Ulrich Mayall (1906–1993) in 1934 and 1939 led Carl Keenan Seyfert (1911–1960; Figure 2) in 1943 to his now famous study of the enhanced activity in the nuclei of six galaxies (or as he called them, 'extragalactic nebulae'). Seyfert, and his predecessors, commented on the similarity with the emission line spectrum of planetary and other gaseous nebulae and noted that the lines are apparently Doppler broadened. There is no evidence that Seyfert ever continued this work, but nevertheless, galaxies containing a stellar nucleus with strong broad emission (including forbidden) lines have become known as 'Seyfert Galaxies'.

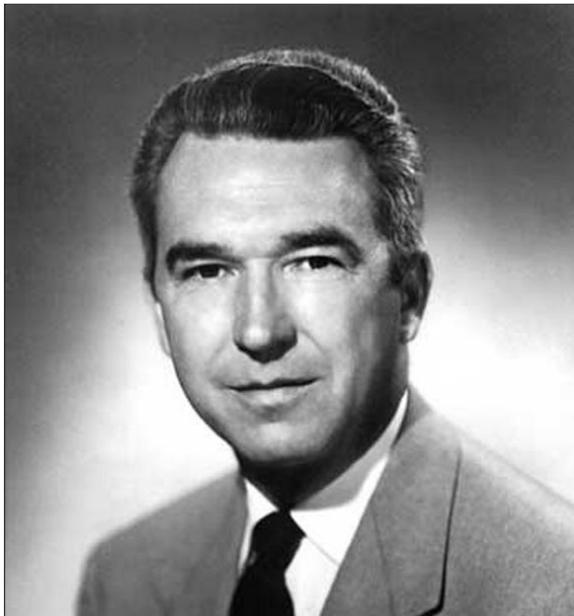

Figure 2: Carl Seyfert (dyer.vanderbilt.edu/about-us/history/).

Curiously, while the SAO/NASA Astrophysics Data System lists 365 citations to Seyfert's 1943 paper, the first one did not appear until a full eight years after Seyfert's 1943 publication during WWII. Even then, not much interest was shown in Seyfert Galaxies until Iosif Samuilovich Shklovsky (1916–1985) drew attention to the possible connection between Seyfert galaxies and radio galaxies (Shklovsky, 1962). Seyfert served three years on the board of Associated Universities Inc. during the critical period when AUI was overseeing the early years of the NRAO, and he had been nominated to serve as the first Director of the NRAO. Unfortunately, in 1960 he died in an automobile accident, but it is interesting to speculate on whether, if Seyfert had lived, his association with the radio astronomers at the NRAO might have led to an earlier appreciation of the relationship between radio emission and nuclear activity.

Even earlier, Sir James Hopwood Jeans (1877–1946) speculated that

> The centres of the nebulae are of the nature of 'singular points,' at which matter is poured into our universe from some other and entirely spatial dimension, so that to a denizen of our universe, they appear as points at which matter is being continually created. (Jeans, 1929: 360).

However, it was Victor Amazaspovich Ambartsumian (1908–1996) who championed the modern ideas that something special was going on in the nuclei of galaxies. At the 1958 Solvay Conference, Ambartsumian (1958: 266) proposed "... a radical change in the conception on the nuclei of galaxies ...", saying that "... apparently we must reject the idea that the nuclei of galaxies is [*sic*] composed of stars alone." He went on to conclude that, "... large masses of prestellar matter are present in nuclei."

In a prescient paper, Fred Hoyle (1915–2001) and William Alfred Fowler (1911–1995) considered

> ... the existence at the very center of galaxies of a stellar-type object of large mass … in which angular momentum is transferred from the central star to a surrounding disk of gas. (Hoyle and Fowler, 1963: 169).

## 2 BEFORE QUASARS

When discrete sources of radio emission were discovered, they were first thought to be due to stars in our Galaxy. Karl Guthe Jansky (1905–1950) and Grote Reber (1911–2002) had shown that the diffuse radio emission was associated with the Milky Way, and since the Milky Way is composed of stars, dust and gas, it seemed natural to suppose that the discrete radio sources were likely connected with stars. Indeed for many years they were called 'radio stars'.

The first hint that at least some radio sources might be extragalactic came from a series of observations made by John Gatenby Bolton (1922

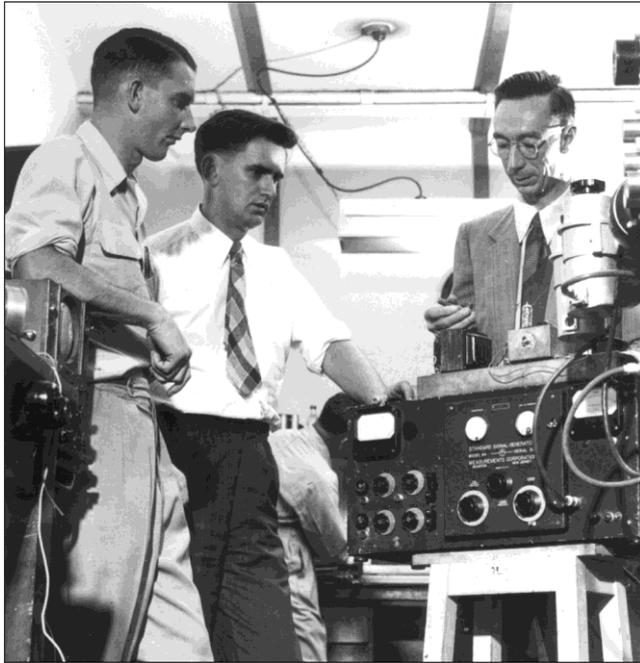 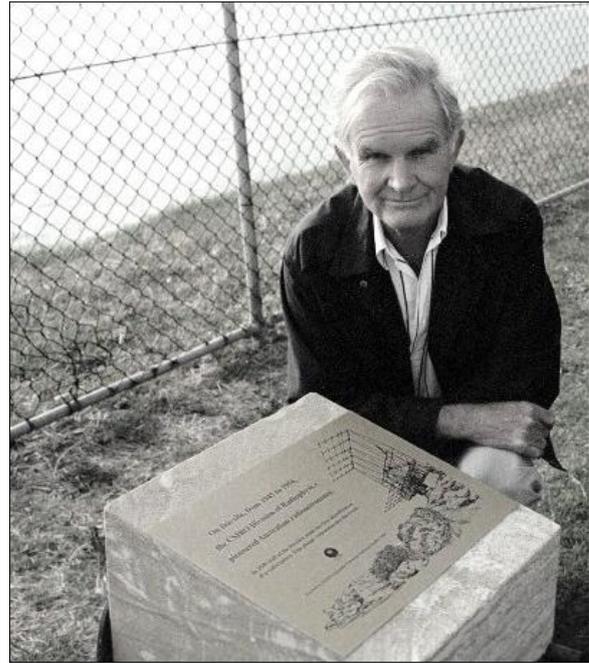

Figure 3: The left hand photograph shows (left to right) John Bolton, Gordon Stanley and the Head of the Radio Astronomy Group at the Division of Radiophysics, Dr Joe Pawsey, at the Radiophysics Laboratory during the late 1940s. The right hand photograph shows Bruce Slee during the unveiling of a commemorative plaque in 1989 at Rodney Reserve, the site of the Division's Dover Heights field Station (both photographs are adapted from originals in the CSIRO's Radio Astronomy Image Archive).

–1993), Gordon James Stanley (1921–2001) and Owen Bruce Slee (b. 1924) (see Figure 3) using cliff interferometers in Australia and New Zealand (see Robertson et al., 2014). After months of painstaking observations, Bolton and his colleagues succeeded in measuring the positions of three strong radio sources with an accuracy of better than half a degree.

For the first time it was possible to associate radio sources with known optical objects. Bolton, Stanley and Slee (1949) identified Taurus A, Centaurus A, and Virgo A with the Crab Nebula, NGC 5128 and M87 respectively. NGC 5128, with its conspicuous dark lane, and M87, with its prominent jet, were well known to astronomers as peculiar galaxies, but their paper, "Positions of Three Discrete Sources of Galactic Radio Frequency Radiation," which was published in *Nature*, mostly discussed the nature of the Crab Nebula. In a few paragraphs near the end of their paper, Bolton, Stanley and Slee commented:

> NGC 5128 and NGC 4486 (M87) have not been resolved into stars, so there is little direct evidence that they are true galaxies. If the identification of the radio sources are [*sic*] accepted, it would indicate that they are [within our own Galaxy].

As implied by the title, Bolton, Stanley and Slee dismissed the extragalactic nature of both Centaurus A and M87. When asked many years later why he did not recognize that he had discovered the first radio galaxies, Bolton (pers. comm., August 1989) responded that he knew they were extragalactic, but that he also realized that the corresponding radio luminosities would be orders of magnitude greater than that of our Galaxy and that he was concerned that in view of their apparent extraordinary luminosities, a conservative *Nature* referee might hold up publication of the paper. However, in a 1989 talk, Bolton (1990) commented that their 1949 paper marked the beginning of extra-galactic radio astronomy. Nevertheless, for the next few years the nature of discrete radio sources remained controversial within the radio astronomy community, and many workers, particularly those at the Cambridge University Cavendish Laboratory, continued to refer to 'radio stars'.

Following the 1954 identification of Cygnus A with a faint galaxy at $z = 0.06$, by Mount Wilson and Palomar astronomers, Wilhelm Heinrich Walter Baade (1893–1960; Figure 4a) and Rudolph Leo Bernard Minkowski (1895–1976; Figure 4b), it became widely appreciated that the high latitude radio sources were in fact very powerful 'radio galaxies', and that the fainter radio sources might be at much larger redshifts, even beyond the limits of the most powerful optical telescopes such as the Palomar 200 inch (Baade and Minkowski, 1954). In a footnote in their paper, Baade and Minkowski noted that Cygnus A previously had been identified with the same galaxy by Bernard Yarnton Mills (1920–2011) and Adin Thomas (1951) and by David Dewhirst (1926–2012) (see Mills and Thomas, 1951; Dewhirst, 1951), but Minkowski confessed that at the time he was not willing to accept the identification with such a faint and distant nebula. Over the next five years, many other radio galaxies were

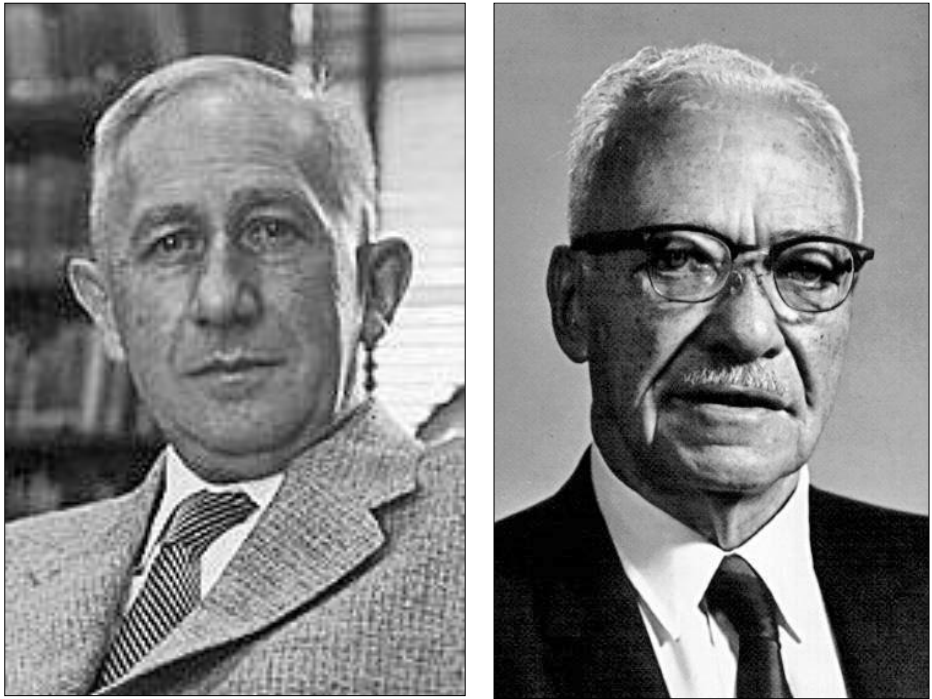

Figure 4a (left): Walter Baade (courtesy: Caltech Archives).
Figure 4b (right): Rudolf Minkowski (courtesy: Astronomical Society of the Pacific).

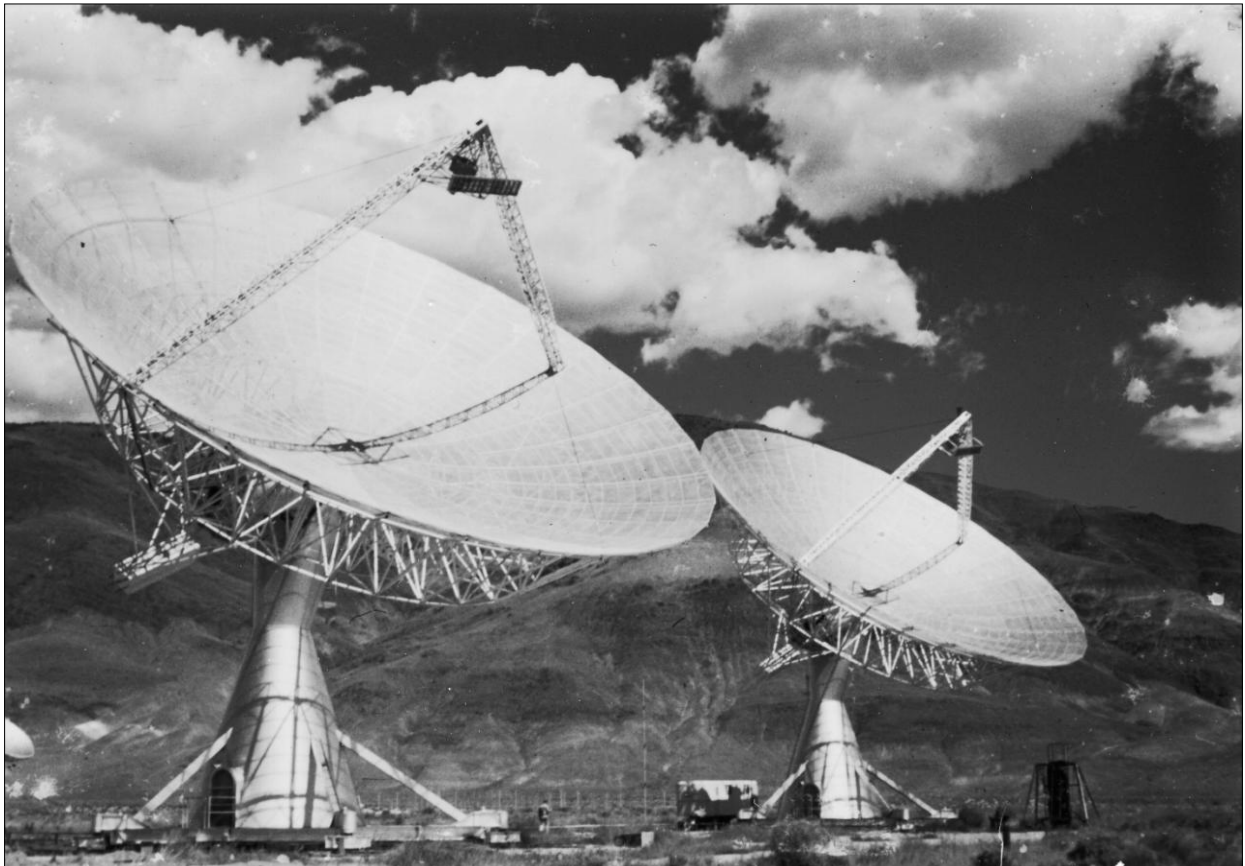

Figure 5: The two-element interferometer at the Owens Valley Radio Observatory (courtesy: Marshall Cohen).

identified, but many others were also misidentified due to inaccurately-measured radio positions.

In 1955, John Bolton came to Caltech from Australia to build a radio telescope specifically designed to accurately measure radio positions and to work with Caltech and Carnegie astronomers to identify and study their optical counterparts. Starting in early 1960, the Caltech OVRO (Figure 5) began to produce hundreds of radio source positions accurate up to ten seconds of arc and leading to new radio galaxy identifications

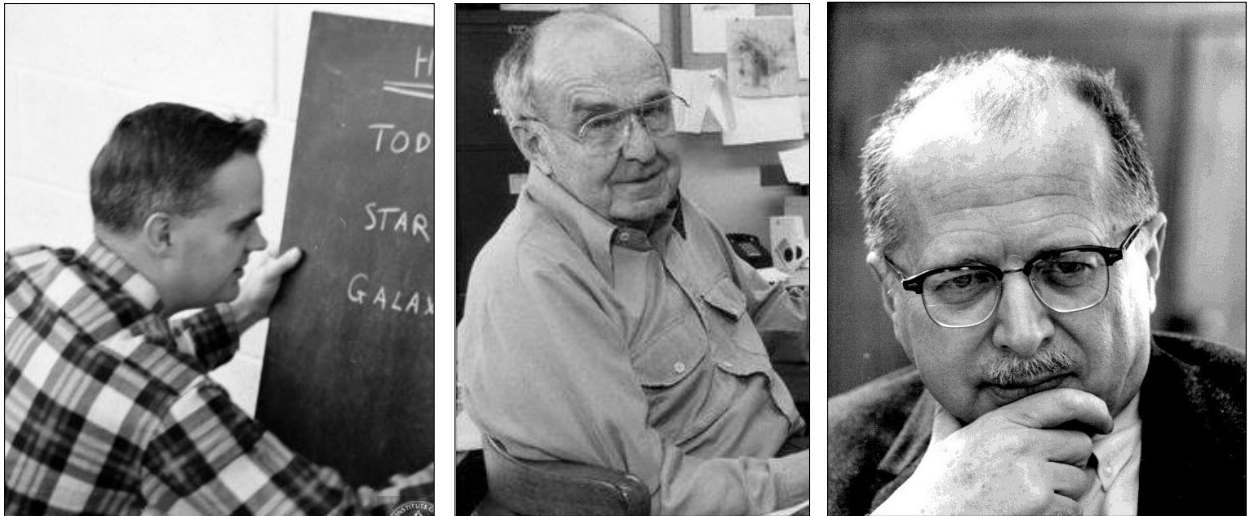

Figure 6a (left): Tom Matthews (courtesy: Caltech Archives).
Figure 6b (center): Alan Sandage (courtesy: Carnegie Observatories).
Figure 6c (right): Jesse Greenstein (courtesy: Caltech Archives).

(Greenstein 1961). Recognizing that radio galaxies were characteristically the brightest galaxy in a cluster (Bolton 1960c), it became clear to many that the search for distant galaxies needed to address the outstanding cosmological problems of the day and therefore should concentrate on galaxies identified with radio sources. Moreover, it was naturally assumed that the smaller radio sources were most likely to be the more distant, so emphasis was given to the smallest radio sources, whose dimensions were determined with the long baseline radio-linked interferometers at Jodrell Bank (Allen et al., 1960; 1962) and with Caltech's OVRO interferometer.

Much of this work was carried out within a collaboration of scientists at Caltech and at the Mount Wilson and Palomar Observatories. John Bolton, Thomas Arnold Matthews (Figure 6a), Alan Theodore Moffet, Richard B. Read and Per E. Maltby (1933–2006) at the OVRO provided accurate radio positions, angular sizes and optical identifications based on inspection of the 48-in Schmidt prints and plates. At the Mt Wilson and Palomar Observatories, Baade, Minkowski, and Allan Sandage (1926–2010; Figure 6b) teamed up with the Caltech radio astronomers to obtain 200-in photographs and spectra. At Caltech, Jesse Greenstein (1909–2002; Figure 6c), Guido Münch (b. 1921) and, after Minkowski's retirement in 1960, Maarten Schmidt, provided spectroscopic follow-up to determine the redshifts of the radio galaxies.

This program had a dramatic success, when, using the 200-in telescope during his last observing session before retiring, Minkowski (1960) found a redshift of 0.46 for the 20.5 magnitude galaxy which was identified by Matthews and Bolton with 3C 295. This made 3C 295 by far the largest known redshift. Although previous to Minkowski's observation, the largest measured spectroscopic redshift was less than 0.2, curiously, an unrelated foreground galaxy located only a few arcminutes from 3C 295 was observed by Minkowski to have a redshift of 0.24, making it the second-largest known redshift at the time. Yet, it would be another 15 years before a galaxy redshift greater than that of 3C 295 would be measured. Interestingly, 3C 295 was targeted not because of any special properties, but only because it was at a high declination, where an accurate declination could be measured with the OVRO interferometer, which until late 1960 had only an East-West base line.

## 3  3C 48: THE FIRST RADIO STAR

By late 1960, it was widely accepted that radio sources located away from the Galactic Plane were powerful distant radio galaxies (e.g., Bolton 1960c). However, Bolton's Caltech colleague, Jesse Greenstein, an acknowledged expert on exotic stars, offered a case of the best Scotch whisky to whoever found the first true radio star. Meanwhile, in the quest to find more distant galaxies, the Caltech identification program concentrated on small diameter sources selected from the early OVRO interferometer observations and from unpublished long-baseline interferometer observations at Jodrell Bank (Allen et al., 1962).

In 1960 Tom Matthews and John Bolton identified 3C 48 with what appeared to be a $16^{th}$ magnitude star. Observations made by Sandage using the 200-in telescope in September 1960 showed a faint red wisp 3″ × 8″ in size. Spectroscopic observations by Sandage, Greenstein and Münch showed multiple emission and absorption lines as well as a strong continuum UV excess, but attempts to identify the lines were inconclusive. On 29 December, Sandage presented a late paper at the $107^{th}$ meeting of the American Astronomical society (AAS) in New York. The listed authors were Matthews, Bolton, Greenstein, Münch and Sandage (1960), which reflected the order of their involvement (Bolton, 1990). But, by this time, Bolton had left Caltech and returned to Australia to oversee the completion and operation of the 64-m Parkes Radio Telescope. Unfortunately, abstracts of late AAS papers were not published at that time, and the only remaining written record of the talk is a news article on the "First True Radio Star" which appeared in the February 1961 issue of *Sky and Telescope* (The first true ..., 1961) and a report in the annual report of the Carnegie Institution of Washington (see Report ..., 1960–1961: 80). Curiously, there is no record of the 107[th] meeting of the AAS at the American Institute of Physics Niels Bohr Library & Archives, although the records of the preceding and succeeding years still exist at the AIP.

*Sky and Telescope* (1961) cautiously reported that

> ... there is a remote possibility that [3C48] may be a distant galaxy of stars; but there is general agreement among the astronomers concerned that it is a relatively nearby star with most peculiar properties.

A few months later, Jesse Greenstein (1961) wrote an article in the Caltech *Engineering and Science* publication announcing "The First True Radio Star."

Subsequent study appeared to confirm the nature of 3C 48 as a true radio star, but with no proper motion as great as 0.05 arcsec/yr. Radio observations indicated angular dimensions less than 0.8 arcsec (Allen et al., 1962) but no measured radio variability (Matthews and Sandage, 1963). Observations by Sandage with the 200-in telescope showed that except for the faint 'wisp', most of the optical light was unresolved, that the color was 'peculiar' and that the optical counterpart varied by at least 0.4 mag over a time scale of months, thus supporting the notion that 3C 48 was a galactic star (Matthews and Sandage, 1963).

Meanwhile, Greenstein (1962) made an exhaustive study of the optical spectrum. After two years of analysis, he submitted a paper to the *Astrophysical Journal* with the title, "The Radio Star 3C48". In this paper, he concluded that 3C 48 was the stellar remains of a supernova, and that the spectrum reflected highly-ionized rare earth elements. The abstract states, "The possibility that the lines might be greatly redshifted forbidden emissions in a very distant galaxy is explored with negative results ...", and the first sentence of the paper states,

> The first spectra of the 16[th] magnitude stellar radio source 3C48 obtained in October 1960 by Sandage were sufficient to show that this object was not an extragalactic nebula of moderate redshift.

Greenstein discussed possible line identifications with several possible redshifts. Although he commented that "... except for 0.367 no red-shift explains the strongest lines of any single ionization ...", he maintained that, "... the case for a large redshift is definitely not proven." Nevertheless, with great prescience, he then went on to point out that if the 3C 48 spectrum is really the red-shifted emission spectrum of a galaxy, then for $\Delta\lambda/\lambda > 1$, Ly-$\alpha$ and other strong UV lines would be shifted into the visible spectrum. When asked years later why he rejected what appeared to be a very important and satisfactory interpretation in terms of a large redshift, Greenstein responded: "I had a reputation for being a radical and was afraid to go out on a limb with such an extreme idea." (Jesse Greenstein, pers. comm., January, 1995).

Subsequently, Matthews and Sandage (1963) succeeded in identifying two additional small-diameter radio sources, 3C 196 and 3C 286, with 'star-like' optical counterparts. But the nature of these 'star-like' counterparts to compact radio sources remained elusive until the investigation of 3C 273 showed them to be distant and unprecedentedly-powerful objects.

## 4  3C 273

3C 273 is the seventh-brightest source in the 3C catalogue. The flux density is comparable to that of 3C 295, so 3C 273 was naturally included in the Caltech program to measure accurate radio positions with the goal of finding optical counterparts for several hundred sources from the 3C catalogue. Typical positional accuracy of the OVRO interferometer was better than 10″ for the stronger sources. Right ascensions were measured by Tom Matthews and others. The declination measurements were part of Richard Read's 1962 Caltech Ph.D. thesis and were submitted for publication in the *Astrophysical Journal* (Read, 1963) in December 1962. The right ascensions were not published until later (Fomalont et al., 1964), but the interferometer positions of many sources, including 3C 273, were available to Caltech astronomers by 1961.

Based on his measured declinations and right ascensions measured by other Caltech radio astronomers, Read (1962) discussed likely identifications of four sources, including 3C 273, with faint galaxies seen only in 200-in plates. Read's declination differed by only 1″ from the currently-accepted position of the quasar. An unpublished contemporaneous right ascension measured by the Caltech radio astronomers was within 2″ of the correct position (see Papers ...). But, the faint galaxy mentioned by Read, which Maarten Schmidt attempted to observe in May 1962, was inexplicably about 1′ west of 3C 273. Interestingly, in his published paper, Read (1963) uses the exact same wording to describe the four identified sources as used in his thesis, but 3C 273 is replaced by 3C 286.

The breakthrough occurred in 1962 as a result of a series of lunar occultations of 3C 273 which were predicted to be observable with the 64-m Parkes Radio Telescope in Australia. Earlier, Cyril Hazard (b. 1928; Figure 7) had used the Jodrell Bank 250-ft Radio Telescope to observe a lunar occultation of the radio source 3C 212. Hazard (1961; 1962) was able to determine the position of 3C 212 with an accuracy of 3″, but his inspection of Palomar Sky Survey plates failed to recognize the 19th magnitude stellar identification which later turned out to be a quasar. Although unresolved by the OVRO interferometer (Moffet, 1962), 3C 273 was known from interferometer measurements at Jodrell Bank (Allen et al., 1962) and Nançay (Lequeux, 1962) to be resolved on longer baselines so it was not on the Caltech list of high-priority small sources that might lead to an identification of a more distant galaxy than 3C 295.

A full occultation of 3C 273, including both immersion and emersion, was predicted to occur on 5 August 1962 and was observed at Parkes at both 136 and 400 MHz. According to Hazard (1977; 1985; Interview ..., 2013) and Bolton (1990), Rudolph Minkowski had a Polaroid copy of a 48-in Schmidt image which indicated the incorrect faint galaxy identification which Schmidt had tried to observe in May of that year. According to Bolton (1990), earlier position measurements made with the Parkes Radio Telescope had established a tentative identification in between a 13th magnitude stellar object and an elongated feature, like an edge-on galaxy, on Minkowski's photograph. However, Bolton claimed that the position accuracy was insufficient to determine whether the radio source was associated with the 'star' or with the elongated feature, so hopefully the occultation would resolve the uncertainty.

The first occultation, on 15 May,[1] when only the immersion was visible, showed diffraction fringes characteristic of a very small source, but was not adequate to derive a position. The 5 August event was more promising as both immersion and emersion were visible, but there was a catch. The emersion of the radio source from behind the Moon was predicted to occur uncomfortably close to the 60° zenith angle limit of the Parkes Radio Telescope (Figure 8). To make sure that the event would not be missed due to an inaccurate position—as had occurred with some earlier predicted occultations—Bolton removed a ladder from the Radio Telescope and ground down part of the gear box in order to ex-tend the length of time the telescope could track. However, as it turned out, the occultation took place at precisely the predicted time, and so the rather drastic alteration of the Radio Telescope's gear box proved unnecessary. But it was a good story which Bolton enjoyed telling and retelling.

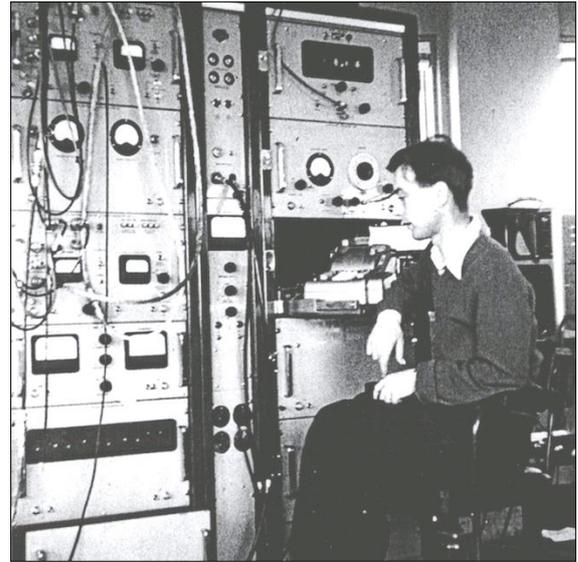

Figure 7: Cyril Hazard observing at Jodrell Bank in the 1950s (courtesy: Miller Goss).

From analysis of the diffraction pattern at 136 and 410 MHz obtained from the immersion (see Figure 9) and emersion of the source from behind the Moon, Hazard et al. (1963) showed that 3C 273 was a complex source consisting of a small flat spectrum component (B) and an elongated steep spectrum component (A). Even before the occultation, Hazard (Interview ..., 1973) was aware of the observations by James Lequeux (1962) using the Nançay interferometer which showed that 3C 273 was a double source with an E-W separation of 14″ between the two components, which was precisely the value derived from the occultation.

On 20 August, just two weeks after the occultation, John Bolton (1962) wrote to Maarten Schmidt at Caltech discussing the Parkes program of radio source identifications and request-ing optical follow-up from Palomar on a radio source located at –28° declination. Somewhat parenthetically, or as an afterthought, he gave the occultation positions of 3C 273 components A and B, and asked Schmidt to pass them along to Tom Matthews, but Bolton made no mention of the obvious optical counterparts in his letter. Moreover, due to a mathematical error by Hazard the positions communicated by Bolton were in error by about 15″. Although Hazard later recounted that Minkowski had tentatively identified 3C 273 with the nebulosity which he assumed to be an edge-on galaxy (Bolton, 1962), in January 1963 Hazard wrote to Schmidt: "I have heard … that *you* have succeeded in identifying the radio source 3C 273 ..." (Hazard, 1963; my italics), so it appears likely that even five months after the occultation neither Hazard nor anyone in Australia was fully convinced of what should have been the obvious association with the bright star that was known to be close to the position of 3C 273. It was probably Tom Matthews (Schmidt, 1963) who provided the precise optical coordinates which showed that the 13th magnitude stellar counterpart was coincident with radio com-ponent B, and that the fainter jet was coincident with radio component A. But, this was only after Schmidt had already observed the spectrum of the stellar object.

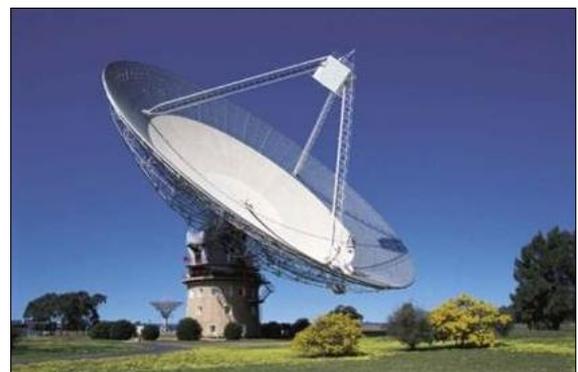

Figure 8: The 64-m Parkes Radio Telescope (photograph: K.I. Kellermann).

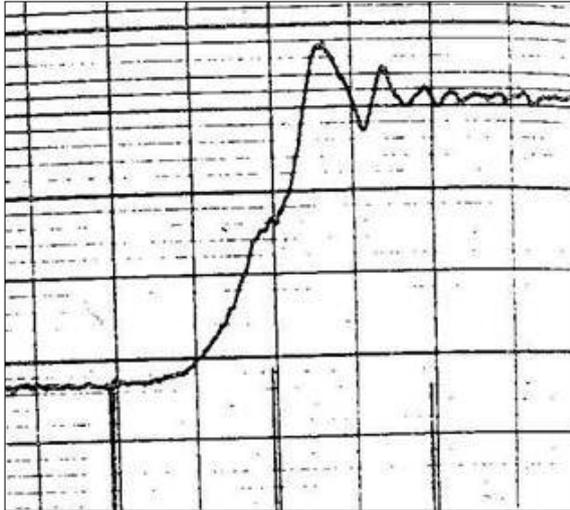

Figure 9: Chart record showing the immersion of the radio source 3C 273 behind the Moon on 5 August 1962 at 410 MHz. Time runs from right to left. The right hand (early) part of the curve shows the Fresnel diffraction pattern as the smaller component B is occulted, followed by the disappearance of the larger component, A (after Hazard et al., 1963; reproduced with permission).

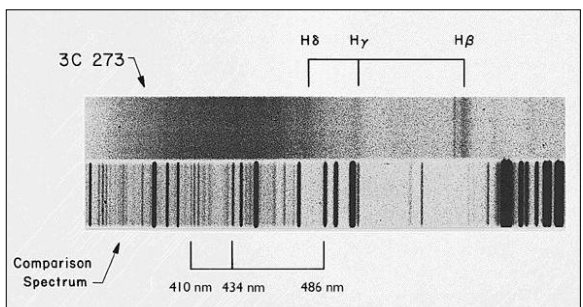

Figure 10: The spectrum of 3C 273 obtained with the 200-in Palomar telescope (courtesy: Maarten Schmidt).

Unfortunately, 3C 273 was too close to the Sun to attempt any spectroscopy until December 1962, when Schmidt used the 200-in telescope to observe the spectrum. Based on the faint optical counterparts that were being associated with other radio sources, Schmidt (1983; 2011) has explained that, like Minkowski, he assumed that the correct optical counterpart was the faint nebulosity. Although at the time there were already three radio sources, 3C 48, 3C 196, and 3C 286, which had been identified with stellar counterparts, Schmidt (ibid.) assumed that that the 13th magnitude stellar object was far too bright to be associated with the radio source. Nevertheless, in order to eliminate the stellar object from further consideration, Schmidt decided to first obtain a spectrum of the 'star'. Having spent most of his 200-in experience observing very faint objects, Schmidt overexposed the spectrum on the first night. Although he obtained a properly-exposed spectrum on 29 December 1962 (Figure 10), it was only on 6 February 1963 that he recognized the simple Balmer series bands of Hβ, Hγ, and Hδ, with a redshift of 0.158 and a corresponding optical luminosity some hundred times brighter than the typical galaxies previously identified with radio sources. Applying that redshift led Schmidt (1963) to recognize the observed band at 3239 Å to be Mg II with a rest wavelength of 2798 Å and another line at a rest wavelength of 5007 Å to be [O III] (Schmidt, 1983). A low-resolution infra-red observation by John Beverley Oke (1928–2004) showed the Hα line shifted to 7599 Å, which confirmed Schmidt's redshift determination (Oke, 1963).

Re-inspection of the spectrum of 3C 48 by Schmidt and Greenstein immediately identified the broad line at 3832 Å to be Mg II redshifted by 0.3679. Using this value of the redshift, the other 3C 48 lines fell into place as [Ne V], [O II], and [Ne III]. Greenstein (1996) recalled that Matthews had earlier suggested "... the possibility of a 37% redshift ..." so the 3C 48 paper was authored by Greenstein and Matthews (1963), but no mention was made of any involvement by John Bolton.[2] However, the high luminosity implied by the observed redshifts along with the small size implied by Sandage's observation of variability was quite remarkable, and was not immediately universally accepted.

The four, now classic, papers by Hazard et al. (1963), Schmidt (1963), Oke (1963), and Greenstein and Matthews (1963) were published as consecutive papers in the 16 March 1963 issue of *Nature*. Whether by error or intentionally,

Hazard's name appeared in *Nature* with a CSIRO[3] Radiophysics Laboratory affiliation, even though he was affiliated with the University of Sydney. At the time relations between the University of Sydney and the Radiophysics Laboratory were already strained, and this incident further exacerbated the existing tensions. Hazard had been invited by John Bolton, Edward George ('Taffy') Bowen (1912–1991) and Joseph Lade Pawsey (1908–1962) to use the Parkes Radio Telescope to observe the occultation. As a non-staff member, Hazard was not familiar with the operation of the Radio Telescope or the instrumentation at Parkes, and so following standard practice for non-Radiophysics observers, CSIRO staff members Albert John Shimmins and Brian Mackey were added to the observing team to provide telescope and instrumental support respectively (Bolton, 1990). Characteristically, Bolton (ibid.) declined to put his name on the paper, stating that he merely "... was just doing his job as Director." Haynes and Haynes (1993) attribute the error to an unintentional mistake on the part of the journal due to the change from a letter format, as submitted, to an article format, as published, although the manuscript submitted by the Radiophysics publications office has the word, 'delete' handwritten in next to Hazard's University of Sydney address.

## 5  3C 48 REVISITED

In 1989 John Bolton gave a talk to the Astronomical Society of Australia in which he reminisced about the period 1955–1960 when he was setting up the radio astronomy program at Caltech. In this talk, Bolton recalled that back in 1960, a full two years before the redshifts of 3C 273 and 3C 48 were announced, he had discussed a possible 3C 48 redshift of 0.37. Bolton (1990) subsequently wrote:

> The best fit I could find for the one broad line and one narrow line which Jesse [Greenstein] had measured were with Mg II λ2798 and [Ne V] λ3426, and a redshift of 0.37.

Since a quarter of a century had passed before his 1989 talk, Bolton's assertion was naturally viewed with skepticism and was emphatically rejected by Greenstein (1996) as a fabrication. However, Fred (later Sir Fred) Hoyle (1915–2001) later recalled that, at the time, John Bolton had told him "I think there's a big redshift in the spectrum." (Hoyle, 1981) Bolton's claim is also supported by a handwritten letter from Bolton to Joe Pawsey dated 16 November, 1960, just one month before Bolton left Caltech for Australia to take charge of the Parkes Radio Telescope, and more than a month before Sandage's AAS paper. Bolton (1960a) wrote:

> I thought we had a star. It is not a star. Measurements on a high dispersion spectrum suggest the lines are those of Neon [V], Argon [III] and [IV] and that the redshift is 0.367. The absolute photographic magnitude is –24 which is two magnitudes greater than anything known. ... I don't know how rare these things are going to be, but one thing is quite clear – we can't afford to dismiss a position in the future because there is nothing but stars.

But, just a month later, on 19 December, Bolton (1960b) wrote to Pawsey from the *SS Orcades*: "The latest news on 3C 48 as I left Caltech was – It is most likely a star." Bolton's letter was mailed from Honolulu where the ship stopped while *en route* to Australia. Apparently, between 16 November and the time he sailed for Australia on 12 December, Bolton had discussed 3C 48 with Ira Sprague Bowen (1898–1973), the then Director of the Mt Wilson and Palomar Observatories who was an expert on spectroscopy. Bowen and Greenstein had inexplicably argued that the dispersion of only three to four angstroms in the calculated rest wavelengths of different lines was too great to accept the large redshift and the corresponding extraordinary absolute magnitude of –24. Indeed, in a paper submitted on 8 December 1962, several months before Schmidt's 3C 273 breakthrough, Matthews and Sandage (1963) had written that they were not able to find "... any plausible combination of red-shifted emission lines." Apparently Matthews had forgotten that earlier he had suggested to Greenstein that 3C 48 might have a 37% redshift. The paper was published in July 1963, with a section discussing "3C 48 as a Galaxy" added in proof after its redshift had been determined by Greenstein and Schmidt.

## 6  QUASARS AND COSMOLOGY

The discovery of quasars with their large redshifts and corresponding unprecedented-large radio and optical luminosities generated a wide range of observational and theoretical investigations as well as a plethora of conferences, particularly the series of Texas Symposia on Relativistic Astrophysics and Cosmology (e.g., Robinson, Schild and Schucking 1965). Motivated by the possibility of extending the Hubble relation to higher redshifts and determining the value of the deceleration constant, $q_0$, for years there was an intense competition to find the highest redshifts. Within two years of the 3C 273 breakthrough, redshifts as high as 2 were reported by Schmidt (1965) and others; but redshifts >3 were not observed until 1973 (Carswell and Strittmatter 1973).

In his classic paper written only five years after he determined the redshift of 3C 273, Schmidt (1968) used a sample of forty quasi-stellar radio sources to derive their luminosity functions and to show that the space density dramatically evolves with cosmic time much in the same manner as powerful radio galaxies. A few years later, Schmidt (1970) extended the study to include optically-selected, e.g., radio quiet, quasars. Now fifty years later, quasar and AGN research have become part of mainstream astronomy with numerous AGN and quasar conferences held each year, along with many books and probably thousands of papers having been published. Supermassive black holes which were first introduced to power quasars (Lynden-Bell, 1969) are now thought to play a major role in galaxy formation and evolution.

## 7 NON-COSMOLOGICAL REDSHIFTS

In view of the apparent extraordinary properties of quasars, a number of well-respected astronomers have argued that the large observed quasar redshifts are intrinsic and not due to cosmological shifts that reflect the expansion of the Universe. The possibility that the observed shifts might be gravitational redshifts was considered very early, but Schmidt and Greenstein (1964) showed that an interpretation in terms of gravitational red-shifts was unrealistic.

Nevertheless, there were a number of observations which appeared to challenge the cosmological interpretation of the large observed redshifts (e.g., Hoyle, 1966; Hoyle and Burbidge, 1966). These included:

(1) *The absence of any redshift-magnitude (Hubble) relation for either the radio or optical data*, now understood in terms of the wide range of apparent quasar luminosities.

(2) *QSO clustering near galaxies*: For many years, Fred Hoyle, Geoffrey Ronald Burbidge (1925–2010), Halton Christian ('Chip') Arp (1927–2013) and others maintained that the density of quasars in the vicinity of galaxies significantly exceeded that found in random fields. Thus, they argued that quasars were ejected from galaxies, but it was difficult to understand the absence of blue shifts in such a model. One not very convincing explanation was that light was emitted only in the opposite direction from the motion in the manner of an exhaust, hence only redshifts were observed. In a variation on this interpretation, James Terrell (1964) suggested that quasars were ejected from the center of our Galaxy and had all passed the Earth, hence we only saw redshifts from the receding objects.

(3) *Distribution of observed redshifts*: Analysis of the distribution of observed quasar redshifts suggested that there were preferred values with peaks near 1.955 (Burbidge and Burbidge, 1967) and at multiples of 0.061 (Burbidge, 1968).

(4) *Radio variability*: The discovery of radio variability and especially rapid inter-day variability in some quasars suggested such correspondingly-small linear dimensions, so if the quasars were at cosmological redshifts the brightness temperature would exceed the inverse Compton limit of $10^{12}$ K (Hoyle and Burbidge, 1966; Kellermann and Pauliny-Toth, 1969). This issue was addressed with the discovery of apparent superluminal motion which is most easily understood as the effect of relativistic beaming which can increase the observed brightness temperature to well above the inverse Compton limit.

(5) *Superluminal motion*: Although relativistic beaming satisfactorily addresses the inverse Compton problem, it was still argued that the observed large angular speed could be more easily understood if quasar redshifts were not cosmological, and the corresponding linear speeds sub-luminal. Indeed, Hoyle, Burbidge and others argued that the relativistic beaming interpretation still required velocities unrealistically close to the speed of light. Still today, the physics of the process by which highly-relativistic motions are attained and maintained remains elusive.

The arguments for non-cosmological redshifts lasted for several decades, and many conferences were held and books written to debate the issues. The apparent anomalies were argued to be the result of *a posteriori* statistics and in the case of redshift distributions, selection effects due to the limited number of strong quasar emission lines that could be observed combined with the narrow range of the observable optical window, and the blocking of certain spectral regions by night-sky lines. The arguments for non-cosmological redshifts only died when the proponents died or at least retired, but from time to time they still surface (e.g., Lopez-Corredoira, 2011).

## 8 INTERLOPERS, BLUE STELLAR OBJECTS, AND QUASI STELLAR GALAXIES

Soon after the discovery of quasars, it was realized that the optical counterparts were unusually blue. This suggested that quasars might be identified by their blue color only, without the need for very precise radio positions. In pursuing radio source identifications with 'blue stellar objects' (BSOs), Sandage (1965) noticed many BSOs that were not coincident with known radio sources, which he called 'interlopers' or 'quasi-stellar galaxies' (or 'QSGs'). In a paper received at the *Astrophysical Journal* on 15 May, 1965, Sandage estimated that the density of interlopers or QSGs was about $10^3$ times greater than that of 3C radio sources. The Editor of the *Astrophysical Journal* was apparently so impressed that he did not send the paper to any referee and by delaying publication was able to rush the paper into publication in the 15 May issue of the *Journal*. Sandage's paper was received with skepticism, no doubt in part generated by what was perceived as privileged treatment of his paper.

Characteristically, Fritz Zwicky (1898–1974; Figure 11) immediately pointed out that

> All of the five quasi-stellar galaxies described individually by Sandage (1965) evidently belong to the subclass of compact galaxies with pure emission spectra previously discovered and described by the present writer. (Zwicky, 1965: 1293).

A few years later, Zwicky was less circumspect and wrote:

> In spite of all these facts being known to him in 1964, Sandage attempted one of the most astounding feats of plagiarism by announcing the existence of a major new component of the Universe: the quasi-stellar galaxies ... Sandage's earthshaking discovery consisted in nothing more than renaming compact galaxies, calling them 'interlopers' and quasistellar galaxies, thus playing the interloper himself. (Zwicky and Zwicky, 1971: xix).

Zwicky (1963) in fact did report at the April 1963 meeting of the AAS that the 'radio stars' are thought to lie at the most luminous end of the sequence of compact galaxies. However, his paper on the same subject was rejected by Subrahmanyan Chandrasekhar (1910–1995), the Editor of the *Astrophysical Journal* editor, who wrote that, "Communications of this character are outside the scope of this journal." (Zwicky and Zwicky, 1971).

Tom Kinman (1965), working at Lick, and Roger Lynds and C. Villere (1965) working at Kitt Peak, each concluded that most of Sandage's BSOs were just that, blue stellar objects, located in our Galaxy and not compact external galaxies. Today we do recognize that there is indeed a population of so called 'radio quiet' quasars, but they are only about an order of magnitude more numerous and three to four orders of magnitude less radio luminous than 'radio loud' quasars.

Interestingly, subsequent investigations later identified several extragalactic radio sources, such as 1219+285, 1514–241 and 2200+420, which had previously been erroneously catalogued as the galactic stars W Comae, AP Lib, and BL Lac respectively.

## 9 NOMENCLATURE

For the lack of any other name Schmidt and Matthews (1964) adopted the term 'quasi stellar radio sources'. This was quite a mouthful, so in a *Physics Today* review article Hong-Yee Chiu (1964) coined the term 'quasar'. This became widely used in oral discussions and in the popular media, but it was not accepted by the *Astrophysical Journal* until, in his 1970 paper on the "Space Distribution and Luminosity Function of Quasars," Schmidt (1970) wrote:

> We use the term "quasar" for the class of objects of starlike appearance (or those containing a dominant starlike component) that exhibit redshifts much larger than those of ordinary stars in the Galaxy. QSOs are quasars selected on the basis of purely optical criteria, while QSSs are quasars selected on both the optical and radio criteria.

Chandrasekhar, the Editor of the *Astrophysical Journal*, responded with a footnote saying:

> The *Astrophysical Journal* has until now not recognized the term "quasar"; and it regrets that it must now concede: Dr. Schmidt feels that, with his precise definition, the term can no longer be ignored.

The term 'quasar' has caught on and is now commonly used in both the popular and professional literature. However, as observations have been extended to cover the entire electromagnetic spectrum, and as improvements in technology have resulted in increasingly-detailed descriptions of both continuum and line spectra, variability, and morphology, quasars have become classified and sub-classified based on their line spectra, as well as their radio, optical, and high-energy spectral distribution. Optical spectroscopy and photometry have defined QSO1s, QSO2s, Broad Absorption Line quasars (BALs), LINERS, and BL Lacs, collectively referred to as QSOs. Radio astronomers have defined Flat Spectrum and Steep Spectrum Radio Quasars, Radio Loud and Radio Quiet Quasars. Radio Quiet quasars have been referred to as In-terlopers, QSGs, and BSOs. Relativistically-beamed quasars are known as blazars; X-ray and gamma-ray observations have defined High and Low Spectral Peaked Quasars. Collectively they are all known as Active Galactic Nuclei (AGN), although the term AGN was originally defined to describe the low luminosity counterpart to powerful quasars.

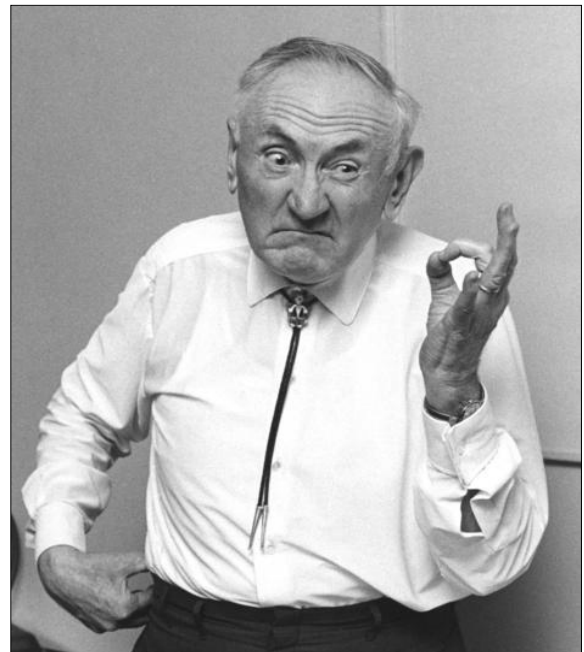

Figure 11: Fritz Zwicky (photograph: Floyd Clark; courtesy: Caltech Archives).

Further background on the early controversies surrounding the discovery of radio galaxies and their implications for cosmology can be found in the excellent books by Edge and Mulkay (1976) and Sullivan III (2009). Shields (1999) and Collin (2006) have given other accounts of the history of quasars and AGN, in particular the subsequent extensive development of the field over the past half century.

## 10  SUMMARY, UNCERTAINTIES, AND SPECULATIONS

Quasars and AGN are now a fundamental part of astrophysics and cosmology. The understand-standing that they are powered by accretion onto a supermassive black hole has had a profound impact on theories of galaxy formation and evolution, and also has motivated extensive research on black hole physics.

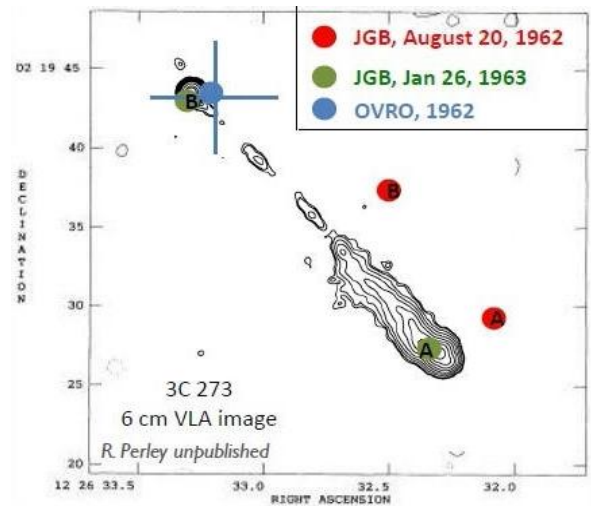

Figure 12: A 6-cm image of 3C 273 obtained with the VLA. Red dots show the location of components A and B as reported to Maarten Schmidt by John Bolton in a letter dated 20 August 1962. Green dots show the location of components A and B as reported by Bolton in his 26 January 1963 letter to Schmidt. The blue point with error bars represents the then-unpublished OVRO interferometer position.

Although both radio and optical astronomers were concentrating on small-diameter radio sources in their quest to locate high-redshift galaxies, the break-through occultation observations of 3C 273 in 1962 were unrelated to the quest for distant galaxies. Indeed, the known relatively-large size of the 3C 273 radio source apparently discouraged further optical investigation until the 1962 series of occultations. Although Bolton (1990) later claimed that the goal of the occultation observation was to determine whether the radio source was associated with the 'star' or the nebulosity, when writing to Schmidt in August 1962 about the occultation, why did he only communicate the radio positions, and uncharacteristically make no mention of the likely obvious optical counterparts. Indeed, he asked Schmidt to "... pass on to Tom [Matthews] the following positions." In his classic paper reporting the redshift, Schmidt (1963) thanked Tom Matthews for drawing his attention to the 13th magnitude 'star', although in a 2012 interview with the author, Matthews denied any involvement in the 3C 273 saga.

It remains unclear why the strong radio source 3C 273 was not identified earlier with its 13th magnitude counterpart. Clearly, Caltech was measuring sufficiently accurate radio source positions as much as two years earlier and had the resources to identify 3C 273 as early as 1961. Although the possibility of identification with apparent stellar objects was already well established with the identifications of the faint optical counterparts 3C 48, 3C 196 and 3C 286, the association with a 13th magnitude stellar object apparently was too extreme to seriously consider. The 19.5 magnitude galaxy first thought to be the optical counterpart of 3C 273 was observed by Schmidt in May 1962, although it was about 1′ west of the true position. Schmidt and the Caltech radio astronomers must have believed that the position error was not more than a few arc seconds, as that kind of accuracy would be required to accept an identification with a barely-detectable galaxy.

An accurate radio position was known among the Caltech radio astronomers perhaps as early as 1961, and certainly prior to Schmidt's May and December 1962 observations made with the 200-in telescope. It seems that somewhere along the line there may have been an error in conveying the OVRO radio position to the Caltech optical astronomers. Ironically, the position conveyed to Schmidt by Bolton on 20 August 1962 which was the basis of the spectrum taken by Schmidt in December was in error by about 15″ due to a wrong calculation by Hazard in determining the time of immersion and emersion during the August occultation. The correct position was not known by Schmidt until he received Bolton's 26 January 1963 letter, more than a month after he obtained the spectrum of the QSO on 29 December, 1962 (see Figure 12).

Thus it appears that the Parkes occultation position played no direct role in the identification of the quasar, which was only recognized as the optical counterpart after Schmidt's 200-in obser-vations showed it to have a peculiar spectrum. Because it was known to be well resolved, 3C 273 had not been on the Caltech list of high-priority sources that might be at high redshift, so it would not have been observed by Schmidt in December were it not for John Bolton's August letter that motivated Schmidt to take the spectrum as the object rose just before the December morning twilight.

Sooner or later, probably sooner, 3C 273 would have been identified on the basis of the OVRO interferometer position, and should have been at least a year or two earlier. Interestingly, it was possible to recognize the 'large' redshift of 3C 273, not because it was large, but because it was so 'small' that the Balmer series was still seen within the small classical optical window. 3C 273 is probably unique in being the only quasar whose spectrum can be so easily determined without prior knowledge of the line identification. It is also somewhat unique in that the radio and optical morphologies are nearly identical, much as in the case of M 87. 3C 273 is the brightest quasar in the radio, IR and optical sky. It is widely believed that the apparent bright radio luminosity is due to relativistic beaming. Does that suggest that the IR and optical emission is also beamed?

The possibility that 3C 48 was a distant galaxy had been discussed by Bolton, Greenstein, Matthews and Bowen at least two full years before the identification of 3C 273. Bolton (1990) later claimed that he and the others apparently rejected the redshift because of a very small 3 to 4 Angstrom discrepancy in the calculated rest wavelengths of the broad emission lines. But, considering the broad nature of the emission lines this seems very unlikely. More probably, they were unwilling to accept the implied huge luminosity, just as Bolton, Stanley and Slee rejected the extragalactic nature of radio galaxies in 1949, and Minkowski later rejected the identification of Cygnus A. It was just too big a step to accept the paradigm-changing radio luminosity until it was forced by Schmidt's interpretation of the 3C 273 spectrum.

It may be relevant that while others were looking for distant galaxies, Bolton may have wanted to believe that he had discovered the first 'radio star' and collect on Greenstein's offer of a case of Scotch, and so 3C 48 played no direct role in the discovery of quasar redshifts. In the end, Greenstein (1963) offered Bolton a beer, but there is no record that Bolton ever accepted. Prior to Schmidt's February 1963 realization of the 3C 273 redshift, Greenstein along with Matthews and Sandage, had submitted separate papers to the *Astrophysical Journal* interpreting the 3C 48 spectrum as a galactic star. After Schmidt's discovery, Greenstein withdrew his paper, while Matthews and Sandage (1963) added a section based on the 0.37 redshift reported by Greenstein and Matthews (1963). Interestingly, on 25 January 1963, just one day prior to his communicating the revised occultation position to Maarten Schmidt, John Bolton gave a lecture to a group of undergraduate students on "Observing Radio Sources". Ron Ekers attended this lecture and his notes show that Bolton stated that "4 genuine radio stars have been identified." (pers. comm., March 2013).[4] At that time only three 'radio stars' were known, 3C 48, 3C 196 and 3C 286, so yet a month after Schmidt had obtained his spectrum Bolton apparently still considered 3C 273 to be among the class of galactic radio stars.

It is curious that John Bolton waited nearly 30 years before making a public claim to have recognized the redshift of 3C 48 two years before the series of *Nature* papers. It is also hard to understand, why Greenstein, apparently having been convinced that 3C 48 was his long sought 'first radio star', used only in-house Caltech publications, *Scientific American* and media releases to announce his claim to the 'first radio star' and did not publish in a refereed journal until after Maarten Schmidt's 1963 breakthrough.

The subsequent search for ever-larger redshifts following the 1963 understanding of the 3C 48 and 3C 273 spectra was highly competitive, particularly among the large optical observatories, but it also contributed to the increasing tensions between Caltech and Carnegie astronomers in Pasadena (see Waluska, 2007).

Currently, the largest-known quasar redshift belongs to ULAS J1120+064 with a redshift of 7.1 (Mortlock et al., 2011). However, it is now gamma-ray bursts and ULIRGs (Ultra Luminous Infra-Red Galaxies), not quasars, that are the most distant and the most luminous objects in the Universe, and gravitational lensing has enabled the detection of a starburst galaxy with a probable redshift of close to 10 (Zheng et al., 2012), well in excess of any known quasar. The controversy over cosmological redshifts was intense and personal and has had a lasting impact on the sociology of astronomy and astronomers.

## 11 NOTES

1. The date is erroneously given as 15 April in the Hazard et al. paper in *Nature*.
2. In his letter Greenstein (1996) claimed that Bolton played no role in the 3C 48 story and that it was Tom Matthews who first suggested that 3C 48 might have a high redshift. In an interview with the author on 28 April 2012, Matthews (Interview ..., 2012) explained that his early suggestion of a large redshift was based simply on the small angular size of the

   radio source, but that once they had the optical identification, he too assumed it was a galactic star. See Section 5.
3. The 'CSIRO' is the Federal Government-funded Commonwealth Scientific and Industrial Research Organisation, and one of it's show-case Divisions was the Division of Radiophysics, based in Sydney, which was a world-leader in radio astronomy.
4. Papers of K.I. Kellermann (NRAO Archives).

## 12 ACKNOWLEDGEMENTS

This paper is based on a paper presented at the Caltech conference "Fifty Years of Quasars" (http://www.astro.caltech.edu/q50/Home.html) and on an earlier version which was published in *The Bulletin of the Astronomical Society of India*. I am indebted to Tom Matthews, Maarten Schmidt, the late Jesse Greenstein, the late Allan Sandage, the late John Bolton, Marshall Cohen, Ron Ekers, John Faulkner, Miller Goss, Woody Sullivan and Jasper Wall for many discussions of the early history of radio galaxies and quasars and for helpful comments on the manuscript. Jesse Greenstein, Allan Sandage, Tom Matthews, Cyril Hazard and Maarten Schmidt shared with me their recollections of the events surrounding the discovery of quasars. Miller Goss and Ron Ekers brought the 1960 and 1961 letters from Bolton to Pawsey to my attention. Maarten Schmidt also kindly shared with me copies of his 1962/1963 correspondence with Hazard and Bolton leading to his 3C 273 red-shift determination. John Faulkner alerted me to the 1981 recollection by Hoyle of Bolton's 1962 claim that 3C 48 had a large redshift. Peter Shaver, Ellen Bouton, Sierra Smith, Richard Porcas, Richard Strom, an anonymous referee, and others provided valuable comments on the manuscript. Shelly Erwin and Loma Karklins gave their generous support in accessing materials at the Caltech Archives.

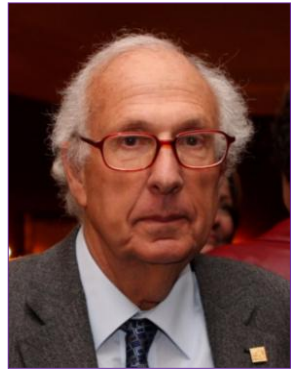

Dr Ken Kellermann is a Senior Scientist at the U.S. National Radio Astronomy Observatory where he does research on radio galaxies, quasars, cosmology and the history of radio astronomy. He did his undergraduate work at MIT and received his Ph.D. in physics and astronomy from Caltech in 1963, after which he spent two years in Australia at the CSIRO Division of Radiophysics. Aside from a two-year stint as a Director at the Max Planck Institute for Radio Astronomy in Bonn, Germany, and various shorter leaves in the Netherlands and back in Australia and at Caltech, he has been at NRAO since 1965. For many years he was active in the International Square Kilometre Array program. He currently works in the NRAO New Initiatives Office where he is involved in the development of new facilities for radio astronomy, and is the current Chair of the IAU Working Group on Historic Radio Astronomy.